\documentclass[aps,pra,letterpaper,twocolumn,superscriptaddress,floatfix,longbibliography]{revtex4-1} 
\usepackage{amsmath,amssymb} 
\usepackage[utf8]{inputenc} 

\usepackage{graphicx} 
\usepackage{hyperref}
\hypersetup{colorlinks=true,citecolor={black},linkcolor={blue},urlcolor={blue}} 
\usepackage{xcolor}
\usepackage{bm}

\newcommand{\mi}{\mathrm{i}}

\newcommand{\ve}{\varepsilon}

\graphicspath{{.}{./figs/}}

\begin{document}

\title{Autonomous chaos of exciton-polariton condensates}

\author{R. Ruiz-Sánchez} 
\affiliation{Instituto de Energías Renovables, Universidad Nacional Autónoma de México, Temixco, Morelos, 62580, Mexico} 

\author{R. Rechtman} 
\affiliation{Instituto de Energías Renovables, Universidad Nacional Autónoma de México, Temixco, Morelos, 62580, Mexico} 

\author{Y. G. Rubo} 
\email{ygr@ier.unam.mx}
\affiliation{Instituto de Energías Renovables, Universidad Nacional Autónoma de México, Temixco, Morelos, 62580, Mexico}

\begin{abstract}
We study the formation of chaos and strange attractors in the order parameter space of a system of two coupled, non-resonantly driven exciton-polariton condensates. 
The typical scenario of bifurcations experienced by the system with increasing external pumping consists of 
(i) formation of $\pi$-synchronised condensates at low pumping, 
(ii) symmetry breaking pitchfork bifurcation leading to unequal occupations of the condensates with non-trivial phase difference between them, 
(iii) loss of the stability of all fixed points in the system resulting in chaotic dynamics, 
(iv) limit cycle dynamics of the order parameter, which ends up in 
(v) in-phase synchronised condensates via the Hopf bifurcation from a limit cycle. 
The chaotic dynamics of the order parameter is evidenced by calculating the maximal Lyapunov exponent. 
The presence of a chaotic domain is studied as a function of polariton-polariton interaction and the Josephson coupling between the condensates. 
At some values of the parameters the bifurcation route is more complex and the strange attractor can coexist with the stable fixed-point lasing. 
We also investigate how the chaotic dynamics is reflected in the light emission spectrum from the microcavity. 
\end{abstract}

\date{\today}
\maketitle

	\section{Introduction\label{sec:intro}}
Spontaneous formation of well-defined polarization has been one of the important features of exciton-polariton (polariton) condensation and lasing in semiconductor microcavities since their discovery more than a decade ago \cite{kasprzak06,balili07,baumberg08,levrat10}. 
Typically, well above the condensation threshold, the condensate is obtained with linear polarization, which is usually interpreted as a result of the minimization of the condensate energy \cite{laussy06,shelykh06}. 
Moreover, the linear polarization is observed to be pinned to a crystallographic X axis, which again can be related to the presence of energy splitting between X and Y linearly polarized cavity modes \cite{kasprzak06,kasprzak07}. 
It is not surprising that for a driven-dissipative system such as the polariton condensate is, these simplistic, energy-related arguments turned out to be rather limited. The polarization behavior observed experimentally is in general more complex, especially in the vicinity of the condensation threshold, and it depends on the way the system is excited and on the fast polarization (spin) dynamics of cavity polaritons \cite{martin01,martin02,lagoudakis02,roumpos09,cerna13}. 
The formation of out-of-equilibrium polariton condensate, or polariton laser, is not governed solely by the energy relaxation, but rather by the whole balance of harvest and decay rates of different single-polariton states together with the polariton-polariton interactions. 

The nontrivial polarization properties of polariton condensation were made especially evident for the condensates spatially detached from the reservoirs of incoherent polaritons, as, e.g., the polariton condensates trapped inside the potential barriers created by the reservoirs \cite{askitopoulos13,cristofolini13}. 
It was shown \cite{ohadi15} that the trapped condensate suffers a bifurcation into a nearly circularly polarized state, stable in a wide pumping range. The formation of circular polarization is even more spectacular than the linear one: the parity symmetry is broken spontaneously, contrary to the explicit XY linear polarization asymmetry of the system. 
For linearly polarized pumping the handedness of the condensate state is chosen randomly, and it can be manipulated by extremely weak electric fields \cite{dreismann16}. The polarization state is even more complex for elliptically polarized pumping, with the domains of polarization inversion and hysteresis of the condensate state as a function of pump intensity \cite{pickup18,valle-inclanredondo19}. 

The properties of single polariton condensate with the polarization degree of freedom are similar to the properties of the pair of condensates separated by a barrier, but with fixed polarization of both (polariton dimer or dyad). The bonding and antibonding single-particle states of dimer are equivalent to the X and Y linearly polarized states, respectively, while the Josephson splitting between them is equivalent to the XY polarization splitting. 
For strong enough Josephson coupling one expects the synchronization of the pair \cite{wouters08,eastham08,voronova15,rahmani16,saito16}, either in the bonding (0 phase difference) or in the antibonding ($\pi$ phase difference) states. 
On the other hand, even for a pair without dissipation, or the Bose-Hubbard dimer, symmetry breaking, self-trapped states are possible \cite{smerzi97,zibold10,bello17}. In these self-trapped solutions, the occupations of two condensates differ strongly, and they are analogous to the strong circular polarization degree in the polarization of a single condensate. 
Interestingly, when the dissipative (or radiative) coupling between the pair is present along with the usual Josephson coupling, two symmetry breaking fixed points can be stable, while two symmetry conserving ones are not, manifesting the weak lasing regime \cite{aleiner12}. In this regime the phase difference between the condensates is non-trivial, between 0 and $\pi$. When all four fixed points of the polariton condensate dimer become unstable the system can exhibit stable limit cycle dynamics, which results in the frequency comb emission from the microcavity~\cite{rayanov15,khan18}. The periodic in time dynamics (or time crystal) can appear in coherently-driven dimer as well~\cite{sarchi08,lledo19,seibold19}. 

The polariton dimer under nonresonant excitation can, in principle, exhibit chaotic dynamics, since it is described by an autonomous system of three nonlinear equations, which is the minimum number of nonlinear equations required to appearance of chaos, the same as for the classical Lorenz \cite{lorenz63} and Rössler \cite{roessler76} systems. 
The Feigenbaum route to chaos through the period doubling of limit cycle orbits has been predicted for actively coupled optical waveguides \cite{alexeeva14}. This case can be considered as a simplistic model for the polariton condensate dimer. Applied to the polariton condensates, this chaotic behavior is expected with decrease of the pumping, so that it takes place at low condensate occupations, well below the threshold. Since this effect precedes the condensate formation, it can hardly be relevant experimentally. 
It was also shown that the non-autonomous chaos of resonantly excited condensates can happen at large condensate occupations \cite{solnyshkov09,gavrilov16}. However, the resonantly excited polariton condensates are not sufficiently stable because of backscattering destabilization and the necessity of fine tuning of the excitation laser. 
In this paper we show that the realistic model of polariton dimer with two condensates being non-resonantly fed by two independent reservoirs, indeed possesses chaotic attractors. It appears at intermediate occupation numbers and can extend to high pumping regime as well, where it can coexist with the synchronised lasing state of the pair.

Understanding of full scenario of bifurcations in the polariton dimer is important, since the dimer is a key element of polariton-condensate networks, which are actively discussed recently for polariton computation and simulation purposes~\cite{berloff17,lagoudakis17,ohadi17,kalinin18}. 
The presence of chaotic dynamics of polariton dimer makes this system very promising for chaos-based applications, including chaos synchronization~\cite{strogatz15book}. As compared to the other optical systems exhibiting deterministic chaos, in particular, to the polarization chaos in vertical-cavity surface-emitting lasers~\cite{sciamanna06,sciamanna09,raddo17,virte18}, the polariton condensate dimer has several benefits, since the important system parameters that control the chaotic dynamics, the Josephson coupling constant and nonlinearities, are easily controlled experimentally.
 
The paper is organized as follows. In Sec.~\ref{sec:model} we describe the theoretical model of the polariton-condensate dimer. 
Sec.~\ref{sec:fixed} analyses the stable-lasing solutions and the domains of their existence and stability. 
In Sec~\ref{sec:lyap} we present the results of calculations of the maximum Lyapunov exponent, which permits to identify the presence of chaotic dynamics. Sec.~\ref{sec:coex} studies coexistence of different lasing states and their probabilities. 
In Sec~\ref{sec:spec} we study the emission spectrum from the microcavity in the chaotic regime. 
Finally, we conclude in Sec.~\ref{sec:conc}.

	\section{The model of a polariton dimer\label{sec:model}}
We consider two polariton condensates, described by the order parameters $\Psi_{+1}$ and $\Psi_{-1}$, 
which obey the driven-dissipative Gross-Pitaevskii equations 
\begin{multline}\label{GPMain}
	\frac{d\Psi_{\pm1}}{dt}=\frac{1}{2}(rN_{\pm1}-\Gamma)\Psi_{\pm1}-\frac{1}{2}(\gamma-\mi\ve)\Psi_{\mp1} \\
	-\frac{\mi}{2}\left[g_1|\Psi_{\pm1}|^2+g_2|\Psi_{\mp1}|^2\right]\Psi_{\pm1},
\end{multline}
coupled to the equations for the densities $N_{\pm1}$ of two non-resonantly excited reservoirs
\begin{equation}\label{Res1}
	\frac{dN_{\pm1}}{dt}=P-\left[\Gamma_R+r|\Psi_{\pm1}|^2\right]N_{\pm1}.
\end{equation}
In these expressions, $\Gamma$ and $\Gamma_R$ are the polariton and reservoir dissipation rates, respectively, $r$ defines the harvest rate of the polaritons into the condensates, and $P$ is the external nonresonant pumping rate. The latter is assumed to be the same for both reservoirs, so that the system of Eqs.\ \eqref{GPMain} and \eqref{Res1} is parity symmetric. 

There is the coherent (Josephson) coupling of two condensates, defined by the parameter $\ve$ and the dissipative coupling between them, given by the parameter $\gamma$. While the former defines the energy splitting of bonding and antibonding states, the presence of the latter implies different lifetimes of these states. Below we consider the typical exciton-polariton condensation case when the antibonding states lives longer due to the destructive interference of the light waves emitted from two centers away from the microcavity \cite{aleiner12}. This difference of lifetimes in grated microcavities can be quite substantial \cite{kim18}. 

The parameters $g_1$ and $g_2$ define the interaction of polaritons in the same and in the opposite centers, respectively. In our model we neglect the interaction of polaritons with the reservoir particles. Apart from avoiding the overload of the model with additional parameters, we intend to consider the case of trapped condensates, which are detached spatially from the reservoirs. This case allows studying unmasked condensate dynamics, which does not suffer from the reservoir noise (see, e.g., Ref.\ \cite{ohadi16} for an example of excitation scheme of such a pair of condensates).  

We introduce the scaled order parameters $\psi_{\pm1}=(r/\Gamma_R)\Psi_{\pm1}$, the reservoir occupations $\tilde{N}_{\pm1}=rN_{\pm1}$, the interaction constants $\alpha_{1,2}=(\Gamma_R/r)g_{1,2}$, and the external pumping $p=rP/\Gamma_R$. Eq.\ \eqref{Res1} is then written as 
\begin{equation}\label{Res2}
	\Gamma_R^{-1}\frac{d\tilde{N}_{\pm1}}{dt}=p-(1+n_{\pm1})\tilde{N}_{\pm1},
\end{equation}
where $n_{\pm1}=|\psi_{\pm1}|^2$ are the scaled occupations of the condensates. 

In what follows, we work in the adiabatic reservoir approximation, commonly used for polariton condensate systems \cite{borgh10,liew15}, which can be applied in the limit of fast reservoir dissipation, $\Gamma_R\gg\Gamma$. In this case, the right-hand-side of Eq.\ \eqref{Res2} is set to zero, and the reservoir occupations are $\tilde{N}_{\pm1}=p/(1+n_{\pm1})$. The order parameters then evolve according to the equations
\begin{multline}\label{GPeqs}
	\frac{d\psi_{\pm1}}{dt}=\frac{1}{2}\left[\frac{p}{(1+n_{\pm1})}-\Gamma\right]\psi_{\pm1}-\frac{1}{2}(\gamma-\mi\ve)\psi_{\mp1} \\
	-\frac{\mi}{2}\left[\alpha_1|\psi_{\pm1}|^2+\alpha_2|\psi_{\mp1}|^2\right]\psi_{\pm1}.
\end{multline}

It is convenient to write $\psi_{\pm1}=\sqrt{n_{\pm1}}\,e^{\mi(\Phi\mp\phi)}$, because the equation for the total phase $\Phi$ is separated from the equations for $n_{\pm1}$ and for the relative phase $2\phi$. The latter variables define the spin vector $\mathbf{S}$ with the components and length
\begin{subequations}\label{SpinDef}
\begin{align}
	& S_x=\sqrt{n_{+1}n_{-1}}\cos(2\phi), \quad S_y=\sqrt{n_{+1}n_{-1}}\sin(2\phi), \\
	& S_z=\frac{1}{2}(n_{+1}-n_{-1}), \qquad S=\frac{1}{2}(n_{+1}+n_{-1}).
\end{align}
\end{subequations}
From \eqref{GPeqs} and \eqref{SpinDef} one can find that the spin components satisfy the equations
\begin{subequations}\label{SpinEqs}
\begin{align}
	& \dot{S}_x=V_x(\mathbf{S})=[u(\mathbf{S})-\Gamma]S_x-{\gamma}S-{\alpha}S_zS_y, \\
	& \dot{S}_y=V_y(\mathbf{S})=[u(\mathbf{S})-\Gamma]S_y+{\ve}S_z+{\alpha}S_zS_x, \\
	& \dot{S}_z=V_z(\mathbf{S})=[u(\mathbf{S})-\Gamma]S_z+v(\mathbf{S})S-{\ve}S_y,
\end{align}
\end{subequations}
where $\alpha=\alpha_1-\alpha_2$ and 
\begin{equation}\label{uvDef}
	u(\mathbf{S})=\frac{(1+S)p}{(1+S)^2-S_z^2}, \quad
	v(\mathbf{S})=-\frac{S_zp}{(1+S)^2-S_z^2}.
\end{equation}
Eqs.\ \eqref{SpinEqs} are the main equations studied in this paper. 
It follows from these equations that the absolute value of the spin evolves as 
$\dot{S}=[u(\mathbf{S})-\Gamma]S+v(\mathbf{S})S_z-{\gamma}S_x$. 
We note that once the evolution of $\mathbf{S}(t)$ is established, one can find the total phase of the condensate by integration of equation
\begin{equation}\label{PhaseDot}
	\dot{\Phi}=-\Omega(\mathbf{S})=-\frac{1}{2}\left[(\alpha_1+\alpha_2)S-\frac{({\ve}SS_x+{\gamma}S_zS_y)}{(S_x^2+S_y^2)}\right].
\end{equation}
In the following sections, we choose the units of time by setting $\Gamma=1$, 
thus measuring all the parameters $\gamma$, $\ve$, and $\alpha_{1,2}$ in the units of $\Gamma$, and time in the units of $\Gamma^{-1}$.

	\section{Fixed lasing states and their stability\label{sec:fixed}}
In general, Eqs.\ \eqref{SpinEqs} possess four fixed points, which are given by the four nontrivial roots of the algebraic system of equations $\mathbf{V}(\mathbf{S})=0$. 
We note that the corresponding order parameters $\psi_{\pm1}$ are not fixed in time, but evolve proportionally to $\exp\{-\mi\Omega(\mathbf{S})t\}$, so that these solutions describe the usual single-mode lasing from the system with the fixed frequency $\Omega(\mathbf{S})$. 
In our model this frequency is counted from the single polariton frequency at the condensation centers. 
There are two symmetry conserving fixed points, $F_s$ and $F_a$, that give equal occupations of the two centers and correspond to symmetric and antisymmetric order parameters, respectively. There are also two symmetry breaking fixed points, $F_+$ and $F_-$, with unequal occupations of the condensation centers, and with $S_z>0$ and $S_z<0$, respectively. 
These fixed point solutions are described in this section in the order of their appearances with increasing pumping $p$.  

\emph{Antisymmetric fixed point.}---The $F_a$ solution appears at the threshold pumping $p_0=\Gamma-\gamma$ from the $S=0$ trivial solution, that becomes unstable for $p>p_0$. The antisymmetric state has $S_y=S_z=0$, $S_x=-S$ and the occupation of the condensates grows linearly with the pumping: $S=(p/p_0)-1$. 
For pumping slightly above $p_0$, the $F_a$ point is the only stable attractor of the system. However, it looses stability with respect to fluctuations of the spin vector in $yz$-plane at the critical spin $S_1$ and the critical pumping $p_1=(1+S_1)p_0$. Standard linear stability analysis shows that the value of $S_1$ can be found as the positive root of equation 
\begin{equation}\label{EqS1}
	\gamma^2+\ve^2+\frac{\gamma(\Gamma-\gamma)S_1}{1+S_1}=\alpha{\ve}S_1.
\end{equation}  

\begin{figure}[t]
	\centering 
	\includegraphics[width=0.48
    \textwidth]{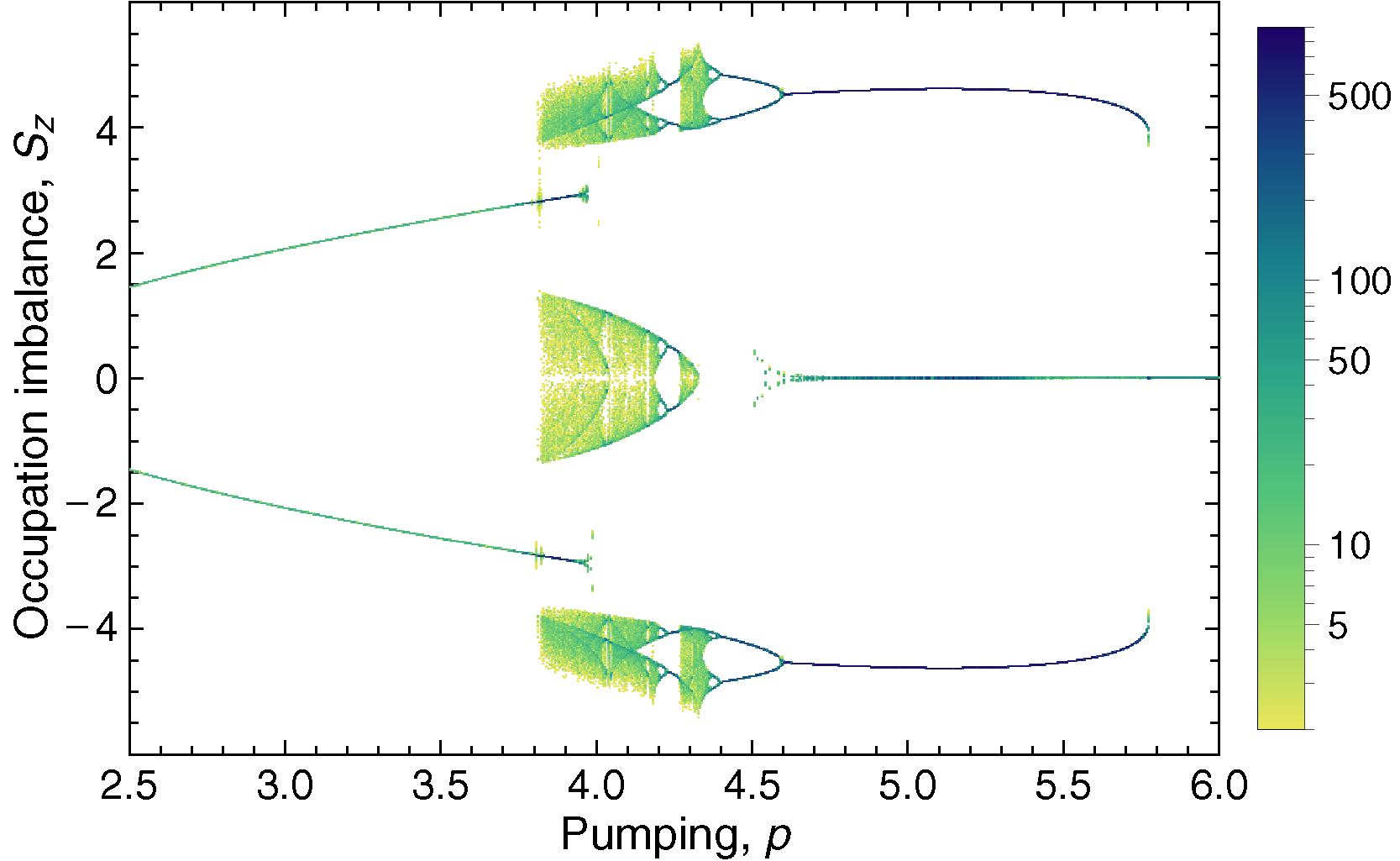} \caption{
	Showing the imbalance of two condensate occupations $S_z$ at the return points of the spin trajectory 
	(points with $dS_z/dt=0$, see text for details).
	The plot has been obtained by collecting the return points for 8 trajectories with random initial conditions at the final stage of 
	evolution between $t=400$ and $t=500$, and for the parameters $\gamma=0.5$, $\ve=2$, $\alpha=0.75$. 
	All parameters are in the units of dissipation rate $\Gamma$.
	}
    \label{fig:1}
\end{figure}
\begin{figure*}[t]
	\centering 
\includegraphics[width=0.48
    \textwidth]{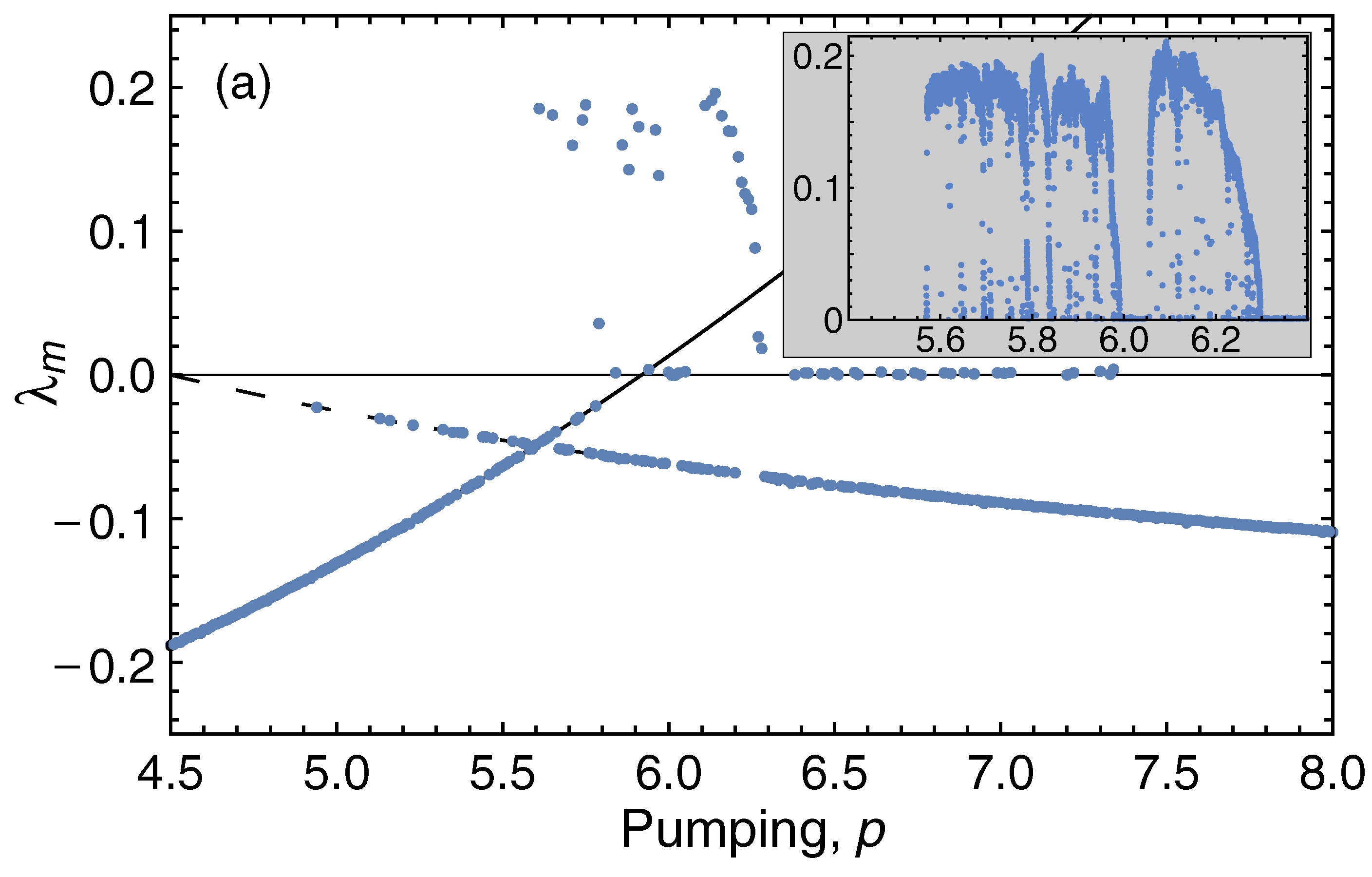}\hspace{0.2in}	
\includegraphics[width=0.48
  \textwidth]{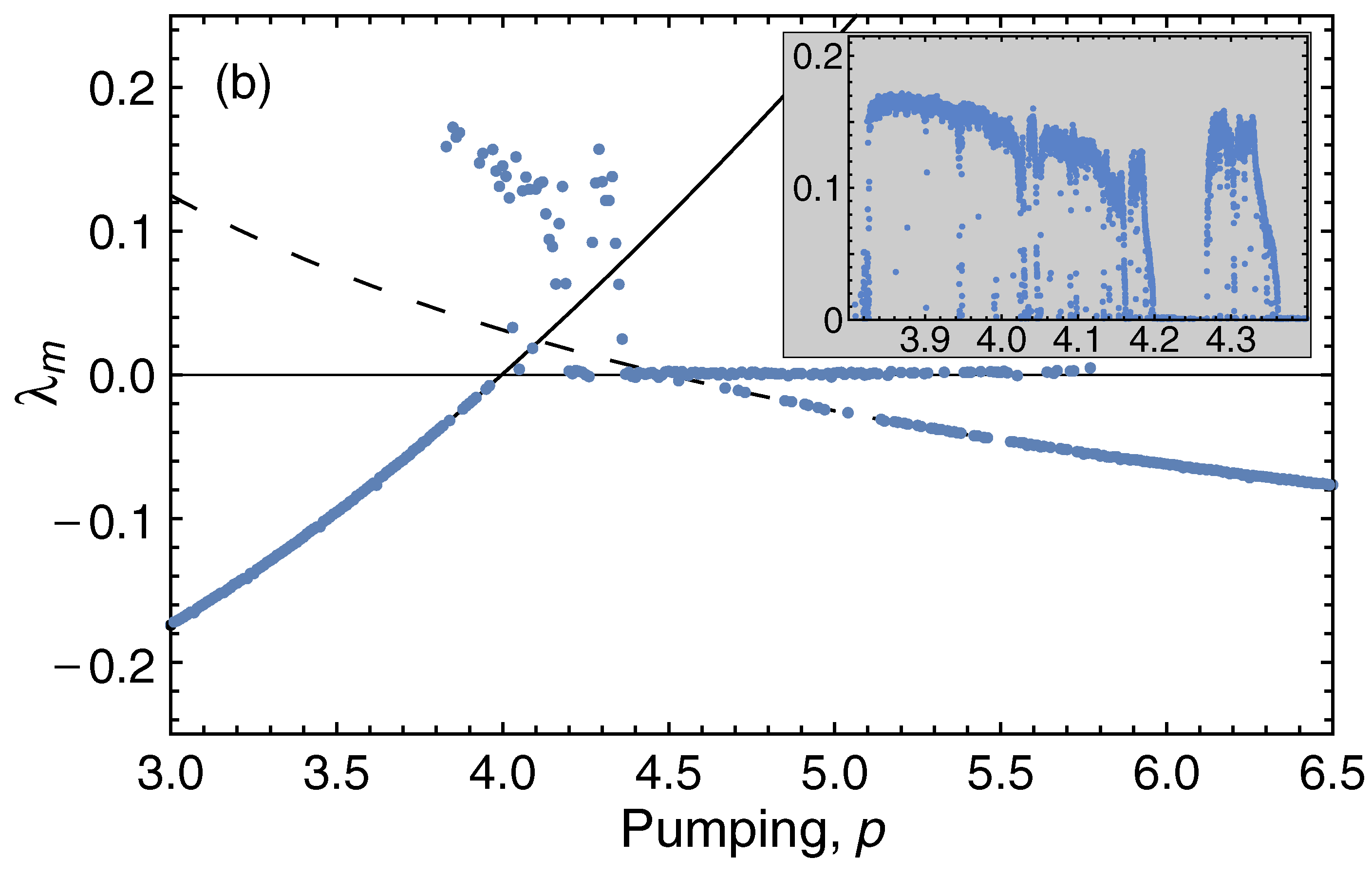}\vspace{0.045in}
\includegraphics[width=0.48
    \textwidth]{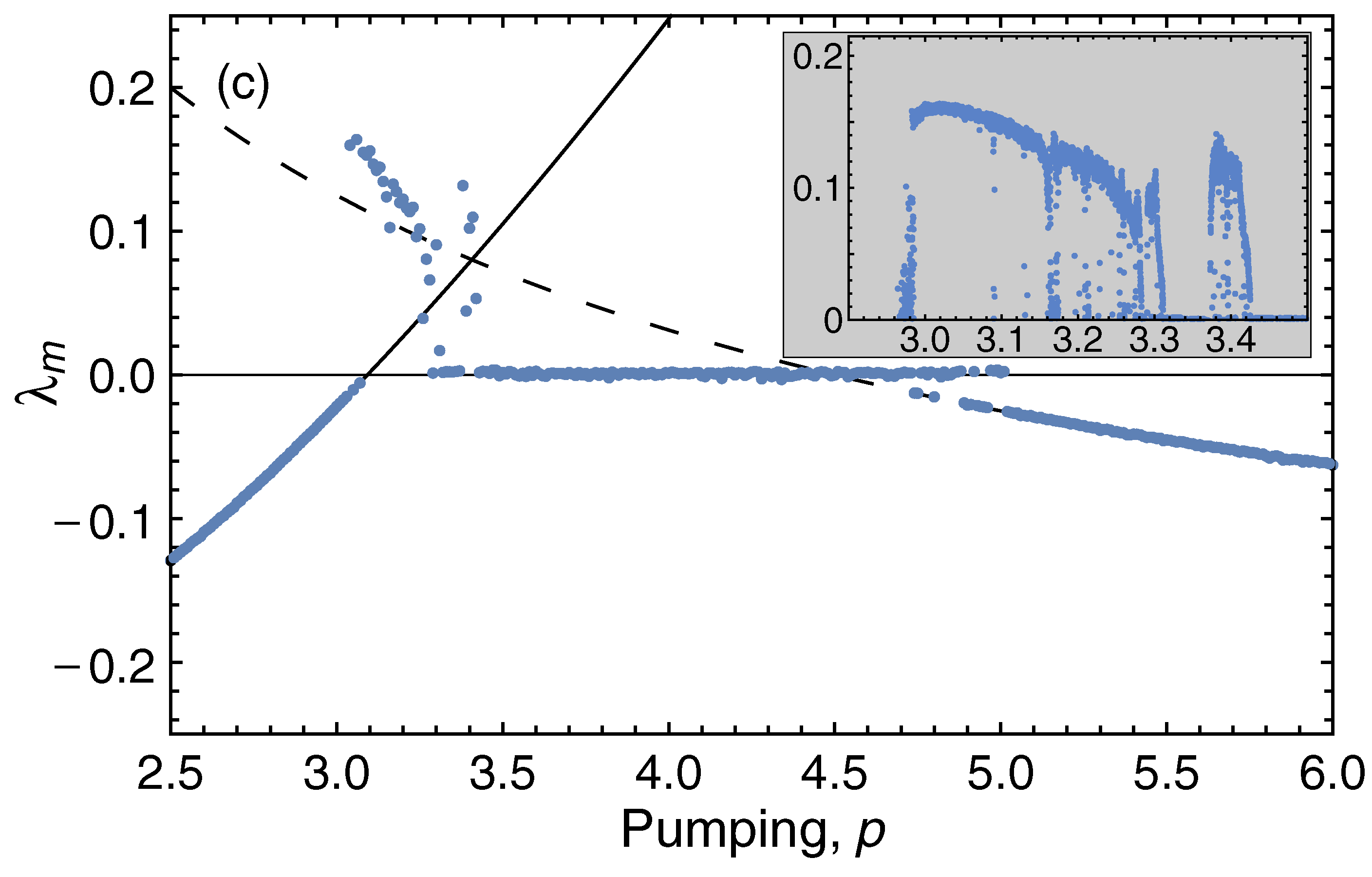} \hspace{0.2in}
\includegraphics[width=0.48
    \textwidth]{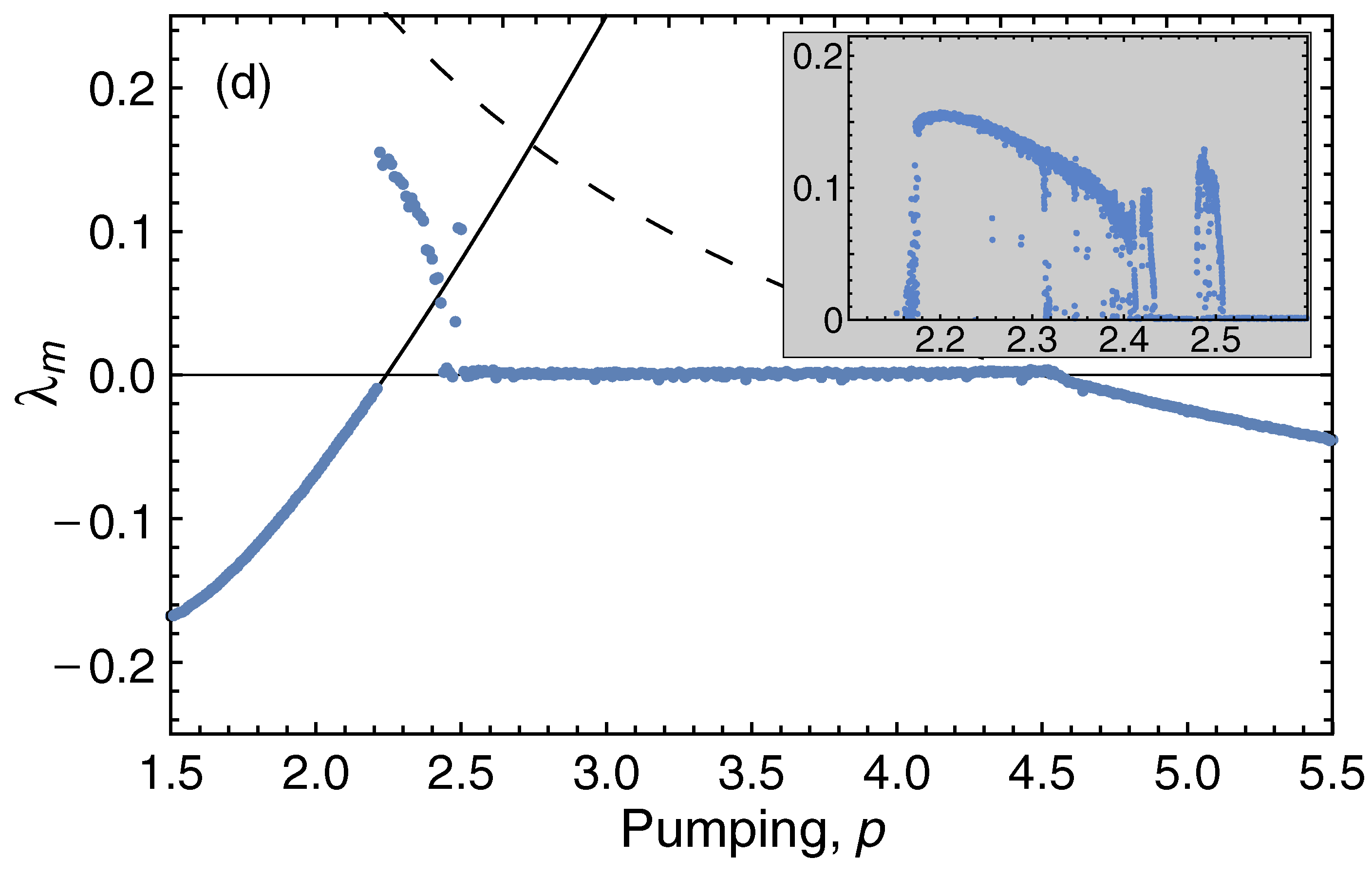} \caption{
The maximum Lyapunov exponent (blue dots) for $\gamma=0.5$, $\ve=2$, and for: 
(a) $\alpha=0.5$, (b) $\alpha=0.75$, (c) $\alpha=1$, and (d) $\alpha=1.5$. 
The MLE for the symmetry breaking fixed points $F_\pm$ is shown by a solid black curve, 
and the MLE for the symmetric fixed point $F_s$ is shown by a dashed black curve.
The time increment in numerical integration is ${\Delta}t=10^{-4}$ and the number of steps is $N=6\times10^{6}$.
The power $p$ steps in 0.01. The insets in the figures show more details of the chaotic domain 
by decreasing the power step and collecting only the points with nonnegative MLE.
	}
    \label{fig:2}
\end{figure*}

\emph{Symmetry breaking fixed points.}---There is a supercritical pitchfork bifurcation at $p=p_1$, the $F_a$ point becomes unstable and two stable fixed points $F_\pm$ are split continuously from it for $p>p_1$. The symmetry between the centers is broken for the new points, and $F_-$ can be obtained from $F_+$ by applying the operations $S_x{\rightarrow}S_x$, $S_y{\rightarrow}-S_y$, and $S_z{\rightarrow}-S_z$, which leave Eqs.\ \eqref{SpinEqs} unchanged. 
Note that the total occupation of two centers is defined by $S$, while $S_z$ defines the occupation imbalance, see Eq.\ (\ref{SpinDef}b). The degree of imbalance $|S_z|/S$ grows monotonically with the pumping strength. In the close vicinity of the bifurcation point $|S_z|/S\propto\sqrt{p-p_1}$. The $F_\pm$ solutions are stable up to some critical pumping $p_2$. 
Although the analytical expression for $p_2$ is rather cumbersome, its value can be found from the analysis of the Lyapunov exponents, which, in this case, coincide with the eigenvalues of the Jacobian matrix ${\partial}V_i/{\partial}S_j$ with $i,j=x,y,z$, calculated at $F_\pm$. 
There are three Lyapunov exponents, one real and always negative, and the other two complex conjugates to each other. 
It is the real part of these pair of complex conjugate Lyapunov exponents that crosses zero at $p_2$, indicating oscillatory behavior of the emerging new attractor.

\emph{Symmetric fixed point.}---The $F_s$ fixed point splits from the trivial fixed point $S=0$ at $p=\Gamma+\gamma$. Similarly to the $F_a$ point, the occupation for this lasing point grows linearly with the pumping, $1+S=p/(\Gamma+\gamma)$. The linear stability analysis shows that this solution is stable with respect to fluctuations in $x$-direction, which are separated from the fluctuations in the $yz$-plane. The Lyapunov exponent for the $x$-direction of the spin is $\lambda_{s1}=-(\Gamma+\gamma)S/(1+S)$. The other two Lyapunov exponents for transverse $yz$-fluctuations are
\begin{equation}\label{Lambdas2}
	\lambda_{s2,3}=\gamma+\frac{1}{2}\lambda_{s1}\pm\sqrt{\frac{1}{4}\lambda_{s1}^2-\ve(\ve+{\alpha}S)}.
\end{equation} 
One can see that when the Josephson splitting $\ve$ and the interaction constant $\alpha$ are not too small, the square root in the above expression is imaginary, and the $F_s$ point becomes stable at $p_3=(\Gamma+\gamma)^2/(\Gamma-\gamma)$. 
For small $\ve$ and $\alpha$, the square root in \eqref{Lambdas2} can be real and this shifts the stability point to higher values of pumping. In what follows, we define the value of the bifurcation pumping point $p_3$ such that the symmetric point is stable for $p>p_3$.  

The typical sequence of bifurcations that the polariton condensate undergoes with increasing pumping is illustrated in Fig.\ \ref{fig:1}. This figure shows the collection of values of $S_z$ at the return points, i.e., the points where $dS_z/dt=0$, as function of pumping strength $p$. For any given pumping, the points have been gathered from eight trajectories obtained from Eqs.\ \eqref{SpinEqs} with random initial conditions at the final stage of evolution. 
For the values of parameters indicated in the caption, the $F_a$ becomes unstable at $p_1\simeq1.98$, and for the small values of pumping shown in the figure one can see two stable fixed points $F_\pm$, with the trajectories randomly ending in one of them. These symmetry-breaking fixed points become unstable at $p_2\simeq4.00$. Close to this value of pumping one can appreciate the presence of another attractor with an unclosed trajectory. 
In the domain of $p$ from 3.8 to 4.2 there is an apparent chaotic behavior, which additionally reappears in a narrow region around $p=4.3$. 
At higher values of pumping there is a limit cycle (LC) motion of the spin, which coexists with the symmetric fixed point, stable for $p>p_3=4.5$.
The diagram suggests that the chaotic attractor appears from the LC dynamics by period doubling bifurcations with decreasing pumping. 
To confirm the presence of autonomous chaos in the polariton condensate described by the system \eqref{SpinEqs}, we study the Lyapunov exponents in the next section.

\begin{figure}[t]
	\centering 
\includegraphics[width=0.49
    \textwidth]{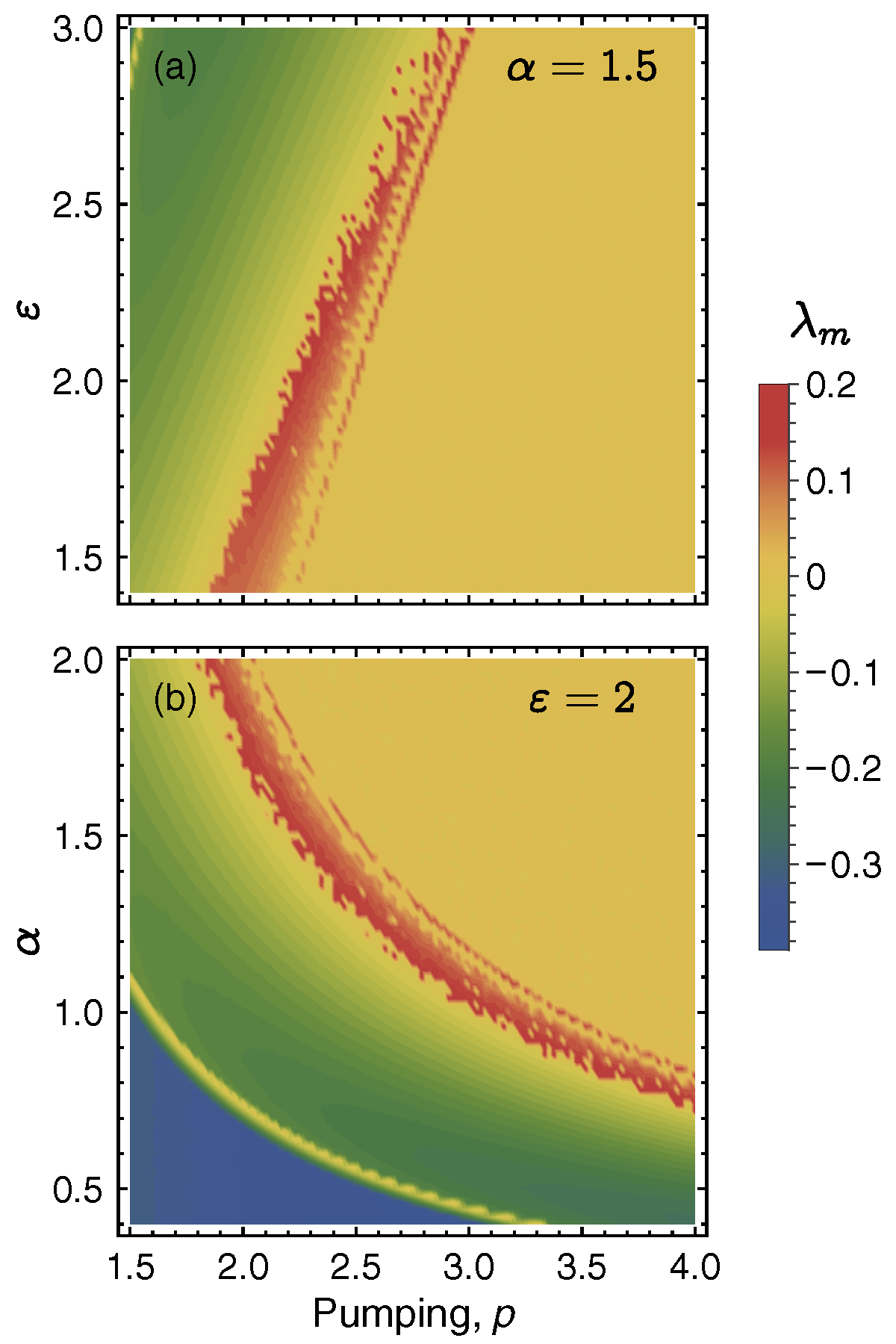}
 \caption{
	The  maximum Lyapunov exponent as a function of the Josephson coupling $\ve$ for $\alpha = 1.5$ in (a), 
	and on the interaction constant $\alpha$ for $\ve=2$ in (b).
	The dissipative coupling is $\gamma=0.5$. 
	The results are obtained with random initial conditions after a transitory evolution. 
	}
    \label{fig:3}
\end{figure}

	\section{The Lyapunov exponents \label{sec:lyap}}
In this section, we study the Lyapunov exponents for the spin dynamics given by Eqs.\ \eqref{SpinEqs}, following the definitions and methods of Refs.\ \cite{benettin80,ott02book,may76,wolf85}. 
We implement the fourth order Runge-Kutta method in the system of Eqs.\ \eqref{SpinEqs} with random initial conditions. 
After transitory evolution to ensure the trajectory to reside on an attractor, we obtain discrete map for the spin vector, 
\begin{equation}\label{SpinMap}
	\mathbf{S}_{k+1}=\mathbf{M}(\mathbf{S}_k)\cdot\mathbf{S}_k, \qquad \mathbf{M}(\mathbf{S}_k)=\mathbf{I}+{\Delta}t\mathbf{J}(\mathbf{S}_k),
\end{equation} 
where $k=0,1,2,\dots,N$ defines the discretisation of the spin trajectory from 0 to $t_\mathrm{max}=N{\Delta}t$, with small time step ${\Delta}t$ and a large number of steps $N$, and $\mathbf{I}$ and $\mathbf{J}$ are the unitary and the Jacobian matrices, respectively. 
The latter is defined by the matrix elements ${\partial}V_i/{\partial}S_j$ with $i,j=x,y,z$ and $\mathbf{V}(\mathbf{S})$ from Eqs.\ \eqref{SpinEqs}. To find the Lyapunov exponents we then use the map \eqref{SpinMap} to study the evolution of the small perturbation vector $\mathbf{s}_k$, renormalizing it in each step. 

First, we discuss the maximum Lyapunov exponent (MLE), which is defined as the maximum real part $\lambda_m$ of the three Lyapunov exponents. This quantity gives a measure of the average rate of divergence or convergence of nearby trajectories in the spin space. Namely, the value $\lambda_m<0$ indicates a trajectory ending in a fixed point, $\lambda_m=0$ characterizes a limit cycle in our case, while $\lambda_m>0$ is the ``smoking gun'' of deterministic chaos, when the trajectory resides in the manifold of a strange attractor \cite{muller95}. 


\begin{figure}[t]
	\centering 
\includegraphics[width=0.48
    \textwidth]{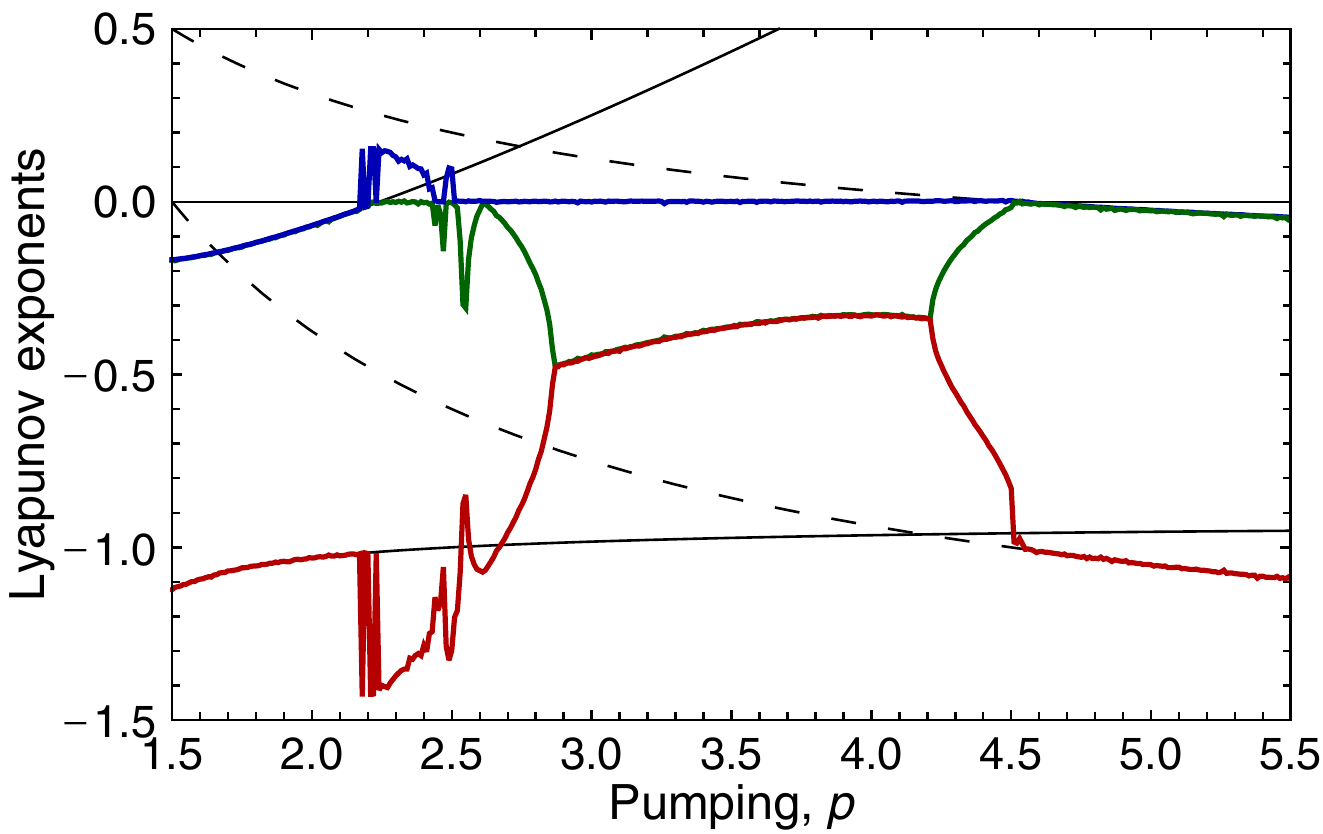}	
 \caption{
	The three Lyapunov exponents (thick lines), calculated for $\gamma=0.5$, $\ve=2$, and $\alpha=1.5$. 
	The thin lines show the Lyapunov exponents for the fixed points, $F_\pm$ (solid) and $F_s$ (dashed).
	The time increment in the Runge-Kutta numerical integration is $\Delta{t}=10^{-4}$. 
	The total integration time after initial transitory evolution is $t_\mathrm{max}=700$.  
	}
    \label{fig:4}
\end{figure}

The results for the MLE are shown in Figs.\ \ref{fig:2}(a-d) for different values of the interaction constant $\alpha$. 
These figures also show the MLE for two relevant fixed-point attractors, $F_\pm$ and $F_s$. Note that for a fixed point, $\lambda_m$ can be also calculated as the maximum real part of the eigenvalues of the Jacobian matrix in this point.
One can appreciate from Figs.\ \ref{fig:2}(a-d) the general scenario of bifurcations in the system. The symmetry breaking fixed point $F_\pm$ is stable for low pumping, while the symmetric fixed point $F_s$ is stable at high pumping $p$. 
In the domain of intermediate pumping, either a strange attractor with $\lambda_m>0$ or a limit cycle with $\lambda_m=0$ are present. 
Since the initial conditions are random, the system randomly chooses an attractor in the case when several stable attractors are present. This results in the dispersion of points in Figs.\ \ref{fig:2}(a-d). 
For weakly interacting polaritons, when both $F_\pm$ and $F_s$ can be stable, the strange attractor can coexist with both, see Fig.\ \ref{fig:2}(a).
The other important feature is the possibility of the coexistence of the stable symmetric fixed point and a limit cycle, see Fig.\ \ref{fig:2}(b) and (c), where the stable LC attractor can extend to rather high values of the pumping $p$. 
These figures suggest that, when the pumping decreases from a high value, the $F_s$ fixed point transforms in to the limit cycle by a Hopf bifurcation, that can be continuous (supercritical), as in Fig.\ \ref{fig:2}(d), or discontinuous (subcritical), as in Figs.\ \ref{fig:2}(a-c). These latter cases result in the coexistence of stable attractors and in the possibility of hysteresis with the pump turning on and off.   

The chaotic behavior of the polariton system depends on both the strength $\alpha$ of the interaction between polaritons and the value of the Josephson coupling between the condensates $\ve$. These dependencies are illustrated in Figs.\ \ref{fig:3}(a,b). The chaotic domain, characterized by the positive values of $\lambda_m$, is shifted to higher pumping with increasing $\ve$ and to lower pumping with increasing $\alpha$.  

During the evolution, the small tangent vector $\mathbf{s}$ is oriented along the direction of the fastest growth. One can use the Gram-Schmidt orthogonalization procedure to establish two other orthogonal directions for the intermediate and the slowest growth. In this way it is possible to calculate the real parts of all three Lyapunov exponents (see Ref.\ \cite{benettin80} for details). In Fig.\ \ref{fig:4} we show the result of these calculations for the same case as in Fig.\ \ref{fig:2}(d), when there is no coexistence of several stable attractors. 
Contrary to known case of the Lorenz system \cite{lorenz63}, where the sum of three exponent is a constant \cite{froyland84}, for our system it is not. Nevertheless, the sum of the Lyapunov exponents is a slowly varying function of the parameters, and the appearance of a chaotic attractor is also manifested in Fig.\ \ref{fig:4} by the additional drop in the value of the smallest exponent.

	\section{Multistability of polariton lasing \label{sec:coex}}
The analysis presented in the previous sections shows that there can be coexistence of several stable attractors in the typical picture of polariton condensation in the vicinity of the threshold. 
Both the symmetry-breaking fixed points and the strange attractor are present in Fig.~\ref{fig:1} just below the critical pumping $p_2\simeq4.0$.
Coexistence of several stable attractors, including fixed points, limit cycles, and/or chaotic attractors, can be observed in Figs.~\ref{fig:2}(a-c). 
These attractors correspond to polariton condensates with different properties, and the possibility of switching between them manifests the multistability of polariton condensation and lasing. 

\begin{figure}[t]
	\centering 
\includegraphics[width=0.48
    \textwidth]{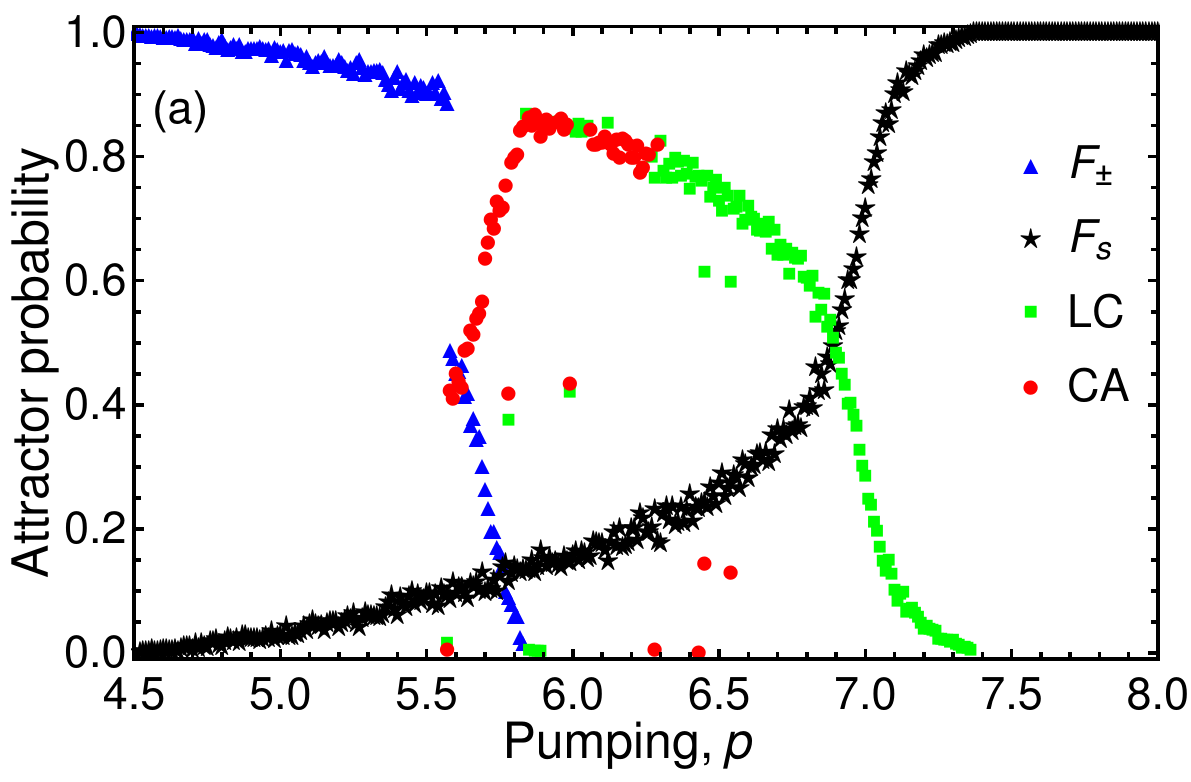}\vspace{0.05in}
\includegraphics[width=0.48
    \textwidth]{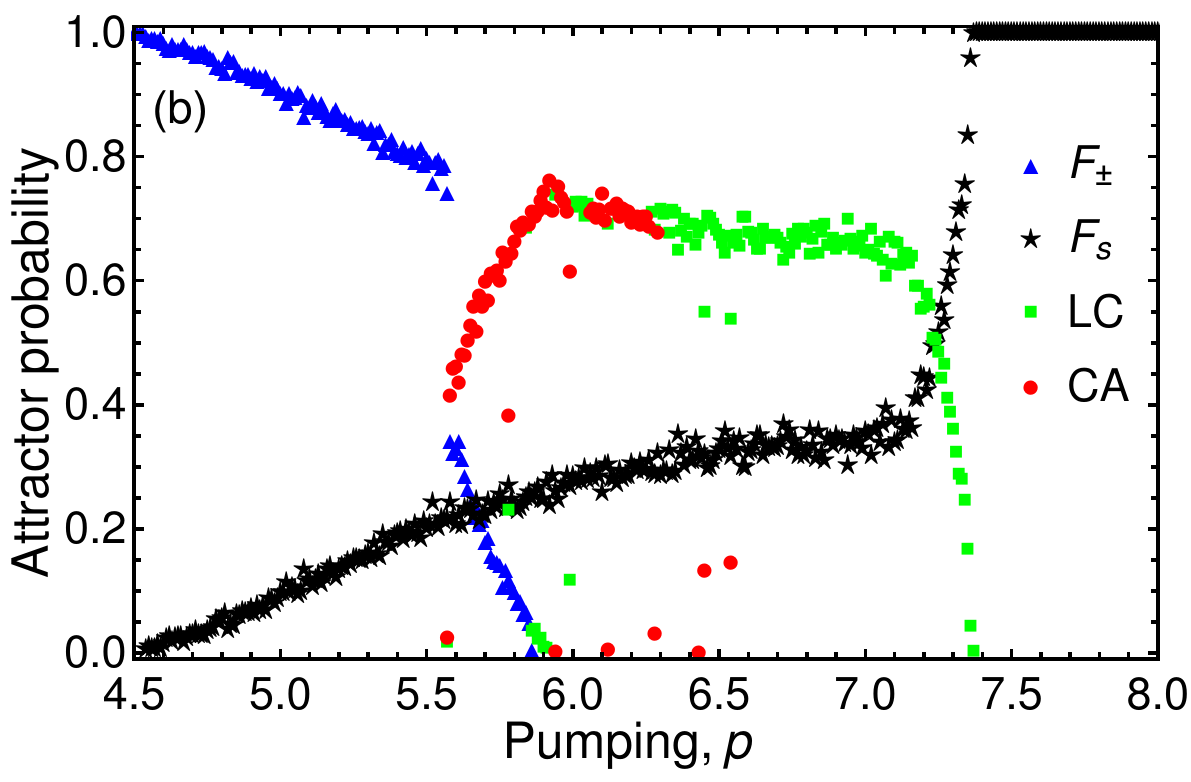}		
 \caption{
	The probabilities of different attractors as functions of pumping, calculated for 1000 runs starting from random initial conditions
	(a) within the large cube in the spin space $-20\le S_{x,y,z}\le 20$, and 
	(b) within the small initial seed $-0.2\le S_{x,y,z}\le0.2$.  
	The parameters are $\gamma=0.5$, $\ve=2$, and $\alpha=0.5$ in the units of $\Gamma$.
	}
    \label{fig:5}
\end{figure}

In the multistability conditions, the formation of one or another attractor (condensate) depends on the initial conditions and the pumping switch. 
The manifold of initial conditions leading to a particular attractor defines the basin of attraction of this final state. 
When the initial conditions are arbitrary, the probabilities of realization of different condensates are given by normalized volumes of corresponding basins of attraction. 
They can be calculated by randomly initializing the system in a big volume of spin space. 
The results of these calculation are shown in Fig.~\ref{fig:5}(a), where the initial conditions resides in a large cube in spin space. 
More relevant for experiments, however, is the case when the condensate grows from a small initial seed, and these probabilities are shown in Fig.~\ref{fig:5}(b). 

Both Figs.~\ref{fig:5} (a) and (b) demonstrate qualitatively similar results. 
The growth of the probability of the symmetric fixed point $F_s$ is continuous and smooth. The presence of this attractor is, therefore, not relevant for the following discussion.  
The other three stable regimes, namely, two symmetry breaking fixed-point lasing $F_\pm$, limit-cycle lasing (LC), and autonomous chaos (AC), clearly compete between themselves. 
Figs.~\ref{fig:5}(a,b) suggest that the $F_\pm$ fixed point looses stability and converts into AC dynamics by a subcritical (type I) bifurcation, so that stable $F_\pm$ and AC coexist in the narrow range of pumping powers $p$, from $p_2^\prime\simeq5.6$ to $p_2\simeq5.9$.

The above scenario is confirmed by studying an adiabatic change of pumping in this domain, that produces characteristic hysteresis behavior. To study this effect we add weak white noise to the right-hand side of Eqs.\ \ref{SpinEqs}. 
When pumping is below $p_2^\prime$ and the condensate is formed in one of the symmetry breaking fixed points, it stays in this state with slow increase of the pumping until $p_2$, where this point becomes unstable and chaotic dynamics appears. 
This transition is accompanied by a drastic growth of the average occupation of the system by about 20\%. 
(Note that the occupation fluctuates substantially in the chaotic regime.) 
When the pumping decreases adiabatically from some $p>p_2$ back, the AC persists until $p_2^\prime<p_2$, where the chaotic dynamic is first transformed into a limit cycle and then into one of the symmetry breaking points by the Hopf bifurcation. 
The $F_+$ and $F_-$ fixed points appear with the same probability in this case, and the bifurcation is also accompanied by the drop in the average occupation of the system. 
This study of hysteresis sheds more light on the nature of chaos in the polariton dimer. 
Each symmetry breaking fixed point produces a limit cycle by the Hopf bifurcation, with one LC residing in $S_z<0$ spin subspace, and the other residing in the $S_z>0$ subspace. When the LC trajectories grow in size, their merging gives rise to chaotic dynamics.

\begin{figure}[t]
	\centering 
\includegraphics[width=0.48
    \textwidth]{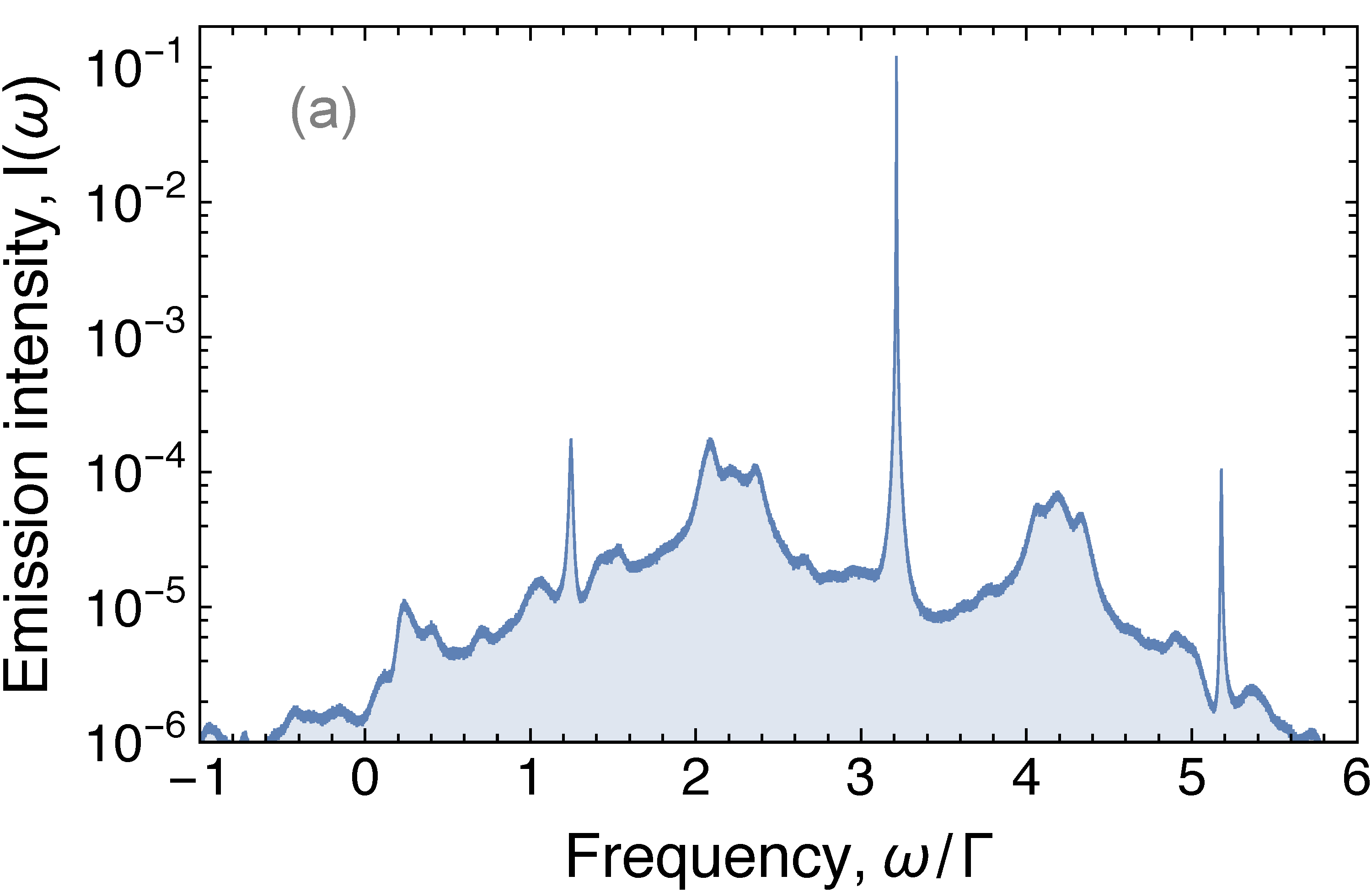}\vspace{0.05in}
\includegraphics[width=0.48
    \textwidth]{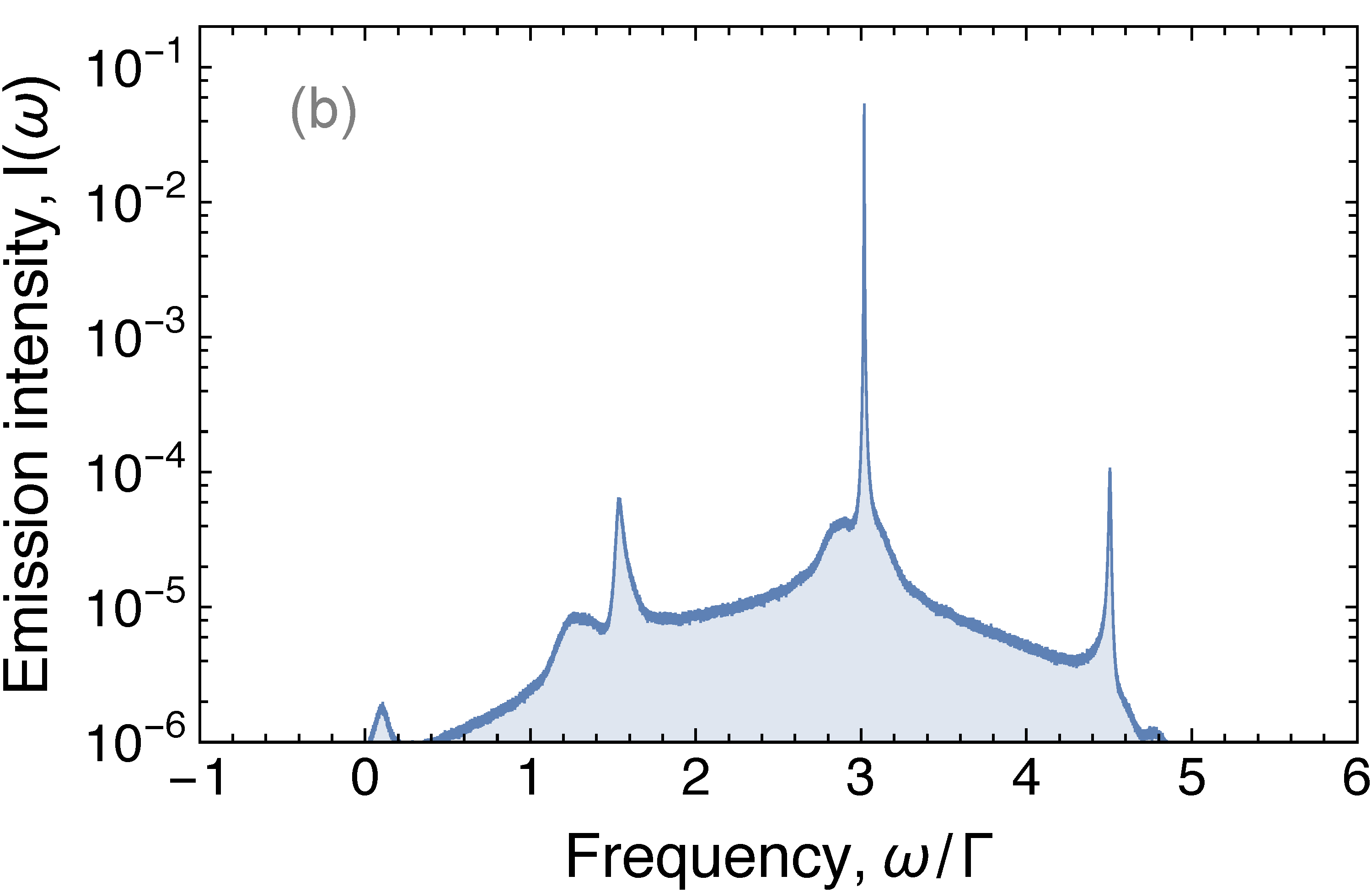}		
 \caption{
	The emission spectrum $I(\omega)$ in the case of chaotic dynamics for $\gamma=0.5$ and $\ve=2$. 
	The interaction strength and pumping are (a) $\alpha=0.5$, $p=6.1$, and (b) $\alpha=1.0$, $p=3.0$. 
	The parameters are in the units of $\Gamma$. See text for more details.
	}
    \label{fig:6}
\end{figure}

	\section{Emission spectrum \label{sec:spec}}
The chaotic dynamics of polariton dimer leads to several interesting features of the emission spectrum from the microcavity. 
Here we calculate the power spectrum $I(\omega)$ using the Fourier transform of the order parameter $\tilde{\psi}_{\pm1}(\omega)$. 
Since the symmetry between the condensation centers is not broken in the chaotic regime, $I(\omega)=|\tilde{\psi}_{\pm1}(\omega)|^2$ is independent of the site index ${\pm1}$. Two examples of the emission spectrum are shown in Figs.~\ref{fig:6}(a,b).  

It turns out that the chaotic dynamic does not lead to just some broad emission. 
The spectrum contains a few pronounced narrow lines that grow on a relatively smooth pedestal. 
To resolve the fine structure of the spectrum, the system of Eqs.~\eqref{GPeqs} has been evolved inside the chaotic attractor for a long time, $t_\mathrm{max}=10^5$, and the trajectory has been discretized with $N=2^{22}$ points to apply the discrete Fourier transform. 
The spectra shown in Figs.~\ref{fig:6}(a,b) have been additionally averaged over 800 random initial conditions in the chaotic attractor manifold.
Note that to calculate the spectrum it is more straightforward to use the equations for the order parameter components \eqref{GPeqs}, not the equations for the spin components \eqref{SpinEqs} together with the equation for the total phase \eqref{PhaseDot}. 
Moreover, it is worth noting that the frequency $\Omega$ defined by Eq.~\eqref{PhaseDot} fluctuates noticeably in chaotic regime and it also contains randomly placed narrow spikes, which appear when the denominator in the second term in \eqref{PhaseDot} becomes close to zero.
In order to highlight the basic features of the chaotic spectrum, we have neither added noise to equations \eqref{GPeqs}, nor included some phenomenological decay.

The main characteristic of the spectra in Figs.~\ref{fig:6}(a,b) is one central narrow line. The position of this line has dynamical origin. 
It is above of the expected blue-shift due to the polariton-polariton interaction. For example, the average site occupation in the case of Figs.~\ref{fig:6}(a) is $\langle{n_{\pm1}}\rangle\simeq7.22$ and the corresponding interaction blue-shift is  $\alpha\langle{n_{\pm1}}\rangle/2=1.8\Gamma$, while the central line is placed at $\omega_0=3.214\Gamma$.
It is important to indicate also that the frequency of the central line is substantially blue-shifted as compared to the position of the lasing from the condensate formed in the symmetric fixed point $F_s$. The $F_s$ lasing takes place with about twice smaller occupation numbers $n_s=[p/(\Gamma+\gamma)]-1\simeq3.07$, and the $F_s$ lasing line is placed at $\omega_s=(\alpha{n_s}-\ve)/2=-0.233\Gamma$.
The shape of the central line is well fitted by the Lorentzian with the full width at half maximum $0.0017\Gamma$. Since the typical values of dissipation rate are $\Gamma\sim0.1\,\mathrm{ps}^{-1}$, the line is quite narrow and the emission can be referred to as lasing in the chaotic regime.

The emission from the pedestal is not negligible. In the case of Figs.~\ref{fig:6}(a), the left wing, the central peak, and the right wing contribute approximately 20\%, 72\%, and 8\% to the total emission, respectively. Interestingly, the spectrum partially resembles the deformed limit cycle emission. The nearest neighbours of the central peak are broad and can be seen as the superposition of three close-placed peaks, while the next-nearest neighbours remain narrow but weak.

	\section{Conclusions \label{sec:conc}}
The formation of polariton lasing in the system of two non-resonantly driven condensates exhibits several nontrivial bifurcations in the vicinity of the threshold. 
When the condensates forming the polariton dimer are coupled both coherently and dissipatively, the bifurcation into stable self-trapped states takes place with increasing pumping. 
The self-trapped states are characterized by broken parity symmetry and different occupations of the condensates. 
Typically, the symmetry-broken fixed point lasing also becomes unstable leading to the development of autonomous chaos. 
The presence of chaos is confirmed by calculating the Lyapunov exponents for a realistic model of a polariton dimer. 
Subsequently, the chaotic dynamics of the system is converted into a limit cycle motion, and at a higher pumping the symmetric synchronized condensate is formed by a Hopf bifurcation from the limit cycle. 
The bifurcations can be both supercritical and subcritical, and the polariton lasing multistability is present in the latter case.       
The frequency spectrum of light emitted from the microcavity in the chaotic regime of polariton condensation is characterized by a few substantially blue-shifted and narrow lines, which grow from a structured pedestal. 

	\section*{Acknowledgements}
We thank Hui Deng and Tim Liew for useful discussions. 
This work was supported in part by CONACYT (Mexico) Grant No.\ 251808, and by PAPIIT-UNAM Grants No.\ IN106320 and IN112719.

%

\end{document}